\documentclass[11pt, a4paper, copyright, goog]{google}

\usepackage[authoryear, sort&compress, round]{natbib}
\usepackage{amsmath}
\usepackage{amssymb}
\usepackage{amsfonts}
\usepackage{bm}
\makeatletter
\@ifpackageloaded{amsthm}{}{\ifdefined\proof\else\usepackage{amsthm}\fi}
\makeatother

\usepackage{graphicx}
\usepackage{booktabs}
\usepackage{multirow}
\usepackage{subcaption}
\usepackage{float}

\usepackage[utf8]{inputenc}
\usepackage[T1]{fontenc}
\usepackage{hyperref}
\usepackage{url}
\usepackage{nicefrac}
\usepackage{microtype}
\usepackage{xcolor}

\hypersetup{pdftitle={Quantifying and Correcting Measurement Error in LLM-Generated Classifications},
  pdfauthor={Yichi Zhang and Ignacio Martinez}}

\newcommand{\eps}{\varepsilon}
\newcommand{\Bin}{\mathrm{Bin}}
\newcommand{\BetaBin}{\mathrm{BetaBin}}
\newcommand{\DE}{\mathrm{DE}}
\newcommand{\Neff}{N_{\mathrm{eff}}}

\newtheorem{theorem}{Theorem}
\newtheorem{lemma}{Lemma}

\newtheorem{theoremA}{Theorem}

\uselogo{}

\title{Quantifying and Correcting Measurement Error in LLM-Generated Classifications: A Beta-Binomial Latent Class Model for Correlated Repeated Judgments}

\correspondingauthor{ychzhang@google.com}

\reportnumber{} 

\renewcommand{\today}{2026-09-30}

\author[1]{Yichi Zhang}
\author[1]{Ignacio Martinez}

\affil[1]{\thepa{}{}}

\begin{abstract}
Large language models are increasingly used to classify text at scale. Because
their outputs are stochastic, it is common to query a model several times per
item and aggregate the results, and latent class models offer a principled way to
convert such repeated ratings into estimates of outcome prevalence and of the
model's error rates. These models assume that repeated ratings of the same item
are conditionally independent given the true label. We examine this assumption
using 52{,}500 judgments from seven open-weight models ranging from 1B to 32B
parameters, collected under three prompt conditions with ten samples per item on
a public benchmark with human labels. On the balanced split of the benchmark,
intraclass correlations lie between 0.50 and 0.93, so ten ratings carry the
information of between 1.07 and 1.81 independent ratings. Under this
correlation, a latent class model with a binomial likelihood is both
miscalibrated and biased. Its nominal 95\% intervals for the false acceptance
rate contained the value computed from human labels in none of 42 settings, and
it underestimated the total error rate in all 42. We propose a beta-binomial
latent class model with a single correlation parameter, establish its
identifiability and that of extensions to covariates, treatment effects and
partially labeled data, and show that on the balanced split it raises the
coverage of nominal 95\% intervals for the prevalence and the false acceptance
rate to 0.90. Recovery of the latent state is governed by the gap between the
model's acceptance rates on truly positive and truly negative items, which equals
Youden's J and predicts recovery almost monotonically (Spearman correlation
0.96). When this gap is small, a labeled subset improves estimates of
prevalence but not the classification of individual items. Simulation studies
designed around the identification conditions support these findings.
\end{abstract}

\providecommand{\keywords}[1]{}
\keywords{Large language models, measurement error, latent class model, overdispersion, LLM as a judge, noisy labels, causal inference}
\paperurl{https://arxiv.org/abs/2510.23874}

\begin{document}

\maketitle

\section{Introduction}
\label{sec:intro}

Large language models (LLMs) are highly effective at automating text
classification and exhibit good alignment with human judgments on many tasks
\citep{bavaresco2024llms}. A fundamental characteristic of these models, their
probabilistic nature, nonetheless presents a challenge. While stochasticity
allows an LLM to generate diverse text, it introduces measurement error in
classification tasks where consistency is required. If an LLM returns the set
\{``dissatisfied'', ``satisfied'', ``dissatisfied'', ``dissatisfied'',
``satisfied''\} for a single customer transcript, the true state of that
transcript is ambiguous. Naive aggregation methods, such as treating a single
round of output as the ground truth or taking the majority vote across several
rounds, fail to characterize the underlying error process, quantify uncertainty,
or accommodate prior information and external covariates.

A common practice is to query the model several times per item and
aggregate. This approach depends on the premise that the repeated samples are
approximately independent draws around a stable propensity for each item, so
that averaging $N$ of them reduces error at the rate familiar from the standard
error of a mean of $N$ independent observations. We test this premise directly
on 52{,}500 judgments from seven open-weight models and find that it does not
hold. In every model and prompt condition we examine, repeated samples are
strongly correlated within items. On the balanced split of the benchmark, ten
draws carry between $1.07$ and $1.81$ independent ratings' worth of information,
and at the median correlation across all settings a standard error computed as
though from ten independent observations is roughly $2.7$ times too small.
Because the correlation is a property of the item rather than of the number of
draws, additional draws do not remove it. Under the exchangeable model fitted at
ten draws, the effective number of independent ratings on the balanced split
cannot exceed two regardless of how many draws are taken, and across all
settings it cannot exceed $3.13$.

A principled way to convert noisy LLM ratings into calibrated estimates is a
latent class model in the tradition of \citet{dawid1979maximum}. Such a model
treats the true label as unobserved, treats the repeated ratings as noisy
measurements of it, and estimates the model's error rates jointly with the
quantity of interest. The standard formulation reduces the ratings of each item
to a count and models that count with a mixture of binomial distributions. This
reduction is valid only when the repeated ratings are conditionally independent
given the true label, which is precisely the assumption that the correlation
within items violates. We show that the consequences are substantial and propose
a model that accounts for them.

This paper makes four contributions.

\begin{enumerate}
\item We quantify the correlation among repeated LLM judgments of the same item
and its consequence for effective sample size, and we show that a latent class
model with a binomial likelihood is both overconfident and systematically biased under
this correlation. Its 95\% intervals for the false acceptance rate cover the
value computed from human labels in 0 of 42 settings, and it underestimates the
total error rate in 42 of 42 (Sections~\ref{sec:independence}
and~\ref{sec:coverage}).

\item We propose a beta-binomial latent class model, which adds a single
parameter for the correlation within items while leaving the error rates and the
prevalence with their original interpretation. We establish its identifiability
(Theorem~\ref{thm:bb}). With the correlation unknown, four ratings per item are
necessary but not sufficient, and five suffice for almost every parameter
value. Identification fails at a correlation of one. We extend the
model to covariates, treatment effects and partially labeled data, and establish
identifiability of each extension. On the balanced split of the benchmark the
proposed model raises the coverage of nominal 95\% intervals for the prevalence
and the false acceptance rate to $0.90$ (Section~\ref{sec:coverage}).

\item We show that recovery of the latent state is governed by the gap between
the model's acceptance rates on truly positive and truly negative items. This
gap equals Youden's $J$ \citep{youden1950index}, its sign is the identifying
restriction of the unsupervised model, and its size predicts recovery almost
monotonically across 42 settings (Section~\ref{sec:separation}). Where the gap
is small, a labeled subset improves estimates of prevalence but not the
classification of individual items (Section~\ref{sec:anchors}).

\item We report simulation studies whose designs follow the identification
conditions of the model, comparing the binomial and beta-binomial likelihoods
with majority voting and single-round classification, and comparing the
semi-supervised model with prediction-powered inference
(Section~\ref{sec:sims}).
\end{enumerate}

The remainder of the paper is organized as follows. Section~\ref{sec:related}
reviews related work. Section~\ref{sec:model} presents the binomial and
beta-binomial latent class models and their identification.
Section~\ref{sec:empirical} reports the empirical study on a public benchmark.
Section~\ref{sec:sims} presents simulation studies, and
Section~\ref{sec:discussion} concludes with a discussion. The collected ratings
and all code are available at \url{https://github.com/google-research/ljer}.

\section{Background and related work}
\label{sec:related}

LLMs have significantly advanced automated text classification in natural
language processing \citep{kostina2025large,liu2023pretrain}, in both business
and scientific settings \citep{bano2024large,ghatora2024sentiment}. However, the
stochastic decoding strategies that enable their generative capabilities
introduce a challenge in classification contexts where consistency and accuracy
are strict requirements \citep{bender2021dangers}. When an LLM is tasked with
classifying an item that has a single true underlying state, its output
variability should be considered a source of
measurement error \citep{atil2024nondeterminism,herrera2025overview,ludwig2025large}.

Addressing this probabilistic nature of LLM outputs is analogous to resolving
conflicts among inconsistent human ratings, for which inter-rater reliability,
truth inference, and annotator competence estimation were developed to infer the
most likely true label \citep{hallgren2012computing,li2019truth,martin2023strong}.
It is also connected to the machine learning literature on learning with noisy
labels \citep{baumann2025large,voutsa2025biased} and to latent class models for
settings without ground truth \citep{cheung2021bayesian,dawid1979maximum}.
\citet{ludwig2025large} propose an applied econometric framework that treats
LLM-generated labels as imperfect surrogates for expensive true measurements,
which relies on a validation sample that is sufficiently large, representative,
and trustworthy. In many applications such a sample is difficult or costly to
obtain \citep{roh2019survey,sambasivan2021everyone,sheng2008get}.

The evaluation of LLM judges is itself well established
\citep{zheng2023judging,yan2026verifybench,li2026verifybench} and typically
reports agreement rates or accuracy against human labels. Our concern is
complementary. Given that a judge errs at some rate and that its repeated
ratings are not independent, we ask how downstream estimates should be formed
and how much uncertainty they carry. \citet{guerdan2025validating} show that
standard validation of judges can obscure genuine indeterminacy in the
underlying ratings, a concern we return to in Section~\ref{sec:limitations}.

Two families of estimators address the downstream problem using a small sample
of gold labels, prediction-powered inference
\citep{angelopoulos2023prediction,angelopoulos2023ppi} and design-based
supervised learning \citep{egami2023using}. Both correct a downstream estimator
using a probability sample of gold labels, and both derive their validity from
the sampling design of that labeled subset rather than from assumptions about
how the model errs. This makes them robust to a poorly behaved annotator but
requires a labeled sample in every application. The model-based approach taken
here differs in three respects: it estimates the model's error rates directly
rather than treating them as nuisance, producing individual posterior
probabilities alongside population estimates; it requires no gold labels 
when the acceptance gap is sufficiently wide (Section~\ref{sec:separation});
and, conversely, it carries greater exposure to model misspecification, 
which is the failure Section~\ref{sec:coverage} documents for
the binomial likelihood. The two approaches are therefore complements. When a
labeled probability sample is available and only a population rate is
required, the design-based estimators rely on fewer assumptions, and
Appendix~\ref{app:sim-semisup} compares the two directly on simulated data.

\section{Methodology}
\label{sec:model}

\subsection{Setup}
\label{sec:setup-model}

Let $c = 1,\dots,C$ index items. Let $D_c \in \{0,1\}$ denote the true and
unobserved binary state of item $c$. In the benchmark study of
Section~\ref{sec:empirical}, $D_c=1$ indicates that the candidate response for
item $c$ agrees with the reference answer. We observe $N_c$ ratings
$R_{c,i} \in \{0,1\}$ from the LLM and write $k_c = \sum_{i=1}^{N_c} R_{c,i}$ for
the number of positive ratings. We define the prevalence
$\theta = \Pr(D_c = 1)$, the \emph{false acceptance rate}
$\eps_0 = \Pr(R_{c,i} = 1 \mid D_c = 0)$, and the \emph{false rejection rate}
$\eps_1 = \Pr(R_{c,i} = 0 \mid D_c = 1)$. The model yields posterior
probabilities $\Pr(D_c = 1 \mid k_c)$ for individual items alongside these
population quantities. Covariates and a treatment indicator can be added, as
described in Section~\ref{sec:bb}.

\subsection{Binomial latent class model}
\label{sec:binom}

When the $N_c$ ratings for an item are conditionally independent given $D_c$,
the count $k_c$ follows a mixture of two binomial distributions,
\begin{equation}
k_c \;\sim\; (1-\theta)\,\Bin(N_c, \eps_0) \;+\; \theta\,\Bin(N_c, 1-\eps_1).
\label{eq:binom}
\end{equation}
A mixture of two binomials with distinct success probabilities and a mixing
weight in $(0,1)$ is identifiable from $N_c \ge 3$ trials up to relabeling of the
two components \citep{teicher1963identifiability,blischke1964estimating}.
Relabeling maps $(\theta,\eps_0,\eps_1)$ to $(1-\theta,\,1-\eps_1,\,1-\eps_0)$
and therefore maps $\eps_0+\eps_1<1$ to $2-(\eps_0+\eps_1)>1$. Restricting the
parameter space to $\eps_0+\eps_1<1$ retains exactly one member of each
relabeled pair.

\begin{theorem}[Identifiability of the binomial latent class model]
\label{thm:binom}
Assume $N_c \ge 3$ conditionally independent ratings are collected for each item
and $\theta \in (0,1)$. The parameters $(\theta, \eps_0, \eps_1)$ are
identifiable if $\eps_0 + \eps_1 < 1$.
\end{theorem}

\noindent The proof is given in Appendix~\ref{app:proof-binom}.

The identifying restriction has a direct interpretation. The LLM accepts a truly
positive item with probability $1-\eps_1$ and a truly negative item with
probability $\eps_0$. The difference between these two acceptance rates is
\begin{equation}
\begin{split}
J &\;=\; (1-\eps_1) - \eps_0 \;=\; 1 - (\eps_0+\eps_1)\\
  &\;=\; \text{sensitivity} + \text{specificity} - 1,
\end{split}
\label{eq:sep}
\end{equation}
which is Youden's $J$ \citep{youden1950index}. The restriction
$\eps_0+\eps_1<1$ therefore states that the LLM accepts positive items more
often than negative items, that is, $J > 0$. We refer to $J$ as the
\emph{acceptance gap}. Section~\ref{sec:separation} shows that its magnitude,
and not only its sign, governs how well the latent state is recovered.

We impose the restriction by writing $v \in (0,1)$ for the total error rate and
$u \in (0,1)$ for the share of it attributable to false acceptance, so that
\begin{equation}
\eps_0 = uv, \qquad \eps_1 = (1-u)v, \qquad \eps_0+\eps_1 = v < 1.
\label{eq:uv}
\end{equation}
This parameterization enforces $J>0$ while allowing either error rate to exceed
$1/2$. The distinction is relevant in practice. The smallest model we evaluate in
Section~\ref{sec:empirical} has $\eps_0 = 0.786$ and a total error rate of
$0.908$ under Arm~A on VerifyBench, so it satisfies $J>0$ although its false acceptance rate exceeds $1/2$.
A constraint that bounds each error rate below $1/2$ separately cannot represent
such a model. Majority voting implicitly relies on that stronger constraint,
since a majority of $N$ ratings recovers the true label only when the model is
correct more often than not for both classes.

\subsection{Beta-binomial latent class model}
\label{sec:bb}

Equation~\eqref{eq:binom} assumes that the ratings of an item are conditionally
independent given $D_c$, and Section~\ref{sec:independence} shows that this
assumption fails for every model we examine. We therefore replace each binomial
component with a beta-binomial component,
\begin{equation}
\begin{split}
k_c \;\sim\;& (1-\theta)\,\BetaBin\big(N_c,\, \eps_0 s,\, (1-\eps_0)s\big)\\
       &+\; \theta\,\BetaBin\big(N_c,\, (1-\eps_1)s,\, \eps_1 s\big),
\end{split}
\label{eq:bb}
\end{equation}
where $s = (1-\rho)/\rho$.
Equivalently, each item has its own acceptance propensity drawn from a beta
distribution with mean $\eps_0$ or $1-\eps_1$ according to its true state, and
its $N_c$ ratings are conditionally independent given that propensity. The
beta-binomial component has the same mean as the binomial component it replaces,
so $\theta$, $\eps_0$ and $\eps_1$ retain their interpretation. The additional
parameter $\rho \in [0,1)$ is the intraclass correlation between two ratings of
the same item, shared by both components, and $\rho = 0$ recovers
Equation~\eqref{eq:binom}. Under Equation~\eqref{eq:bb} the variance of the
acceptance fraction $k_c/N$ within a component exceeds its binomial value by the
factor
\begin{equation}
\DE \;=\; 1 + (N-1)\rho,
\label{eq:de}
\end{equation}
which is the design effect of cluster sampling, and the effective number of
independent ratings is $\Neff = N/\DE$. The parameter $\rho$ can therefore be
compared directly with the correlation measured in Section~\ref{sec:independence}.

\begin{theorem}[Identifiability of the beta-binomial latent class model]
\label{thm:bb}
Assume each item receives $N$ ratings, $\theta \in (0,1)$,
$\eps_0, \eps_1 \in (0,1)$, $\eps_0 + \eps_1 < 1$, and $0 \le \rho < 1$, where
$\rho = 0$ denotes the binomial model.
\begin{enumerate}
\item[(i)] If $\rho$ is known, $(\theta, \eps_0, \eps_1)$ are identifiable for
$N \ge 3$.
\item[(ii)] If $\rho$ is unknown, $N \ge 4$ is necessary for identifiability of
$(\theta, \eps_0, \eps_1, \rho)$ but not sufficient. There is a set of parameter
values of positive Lebesgue measure on which two distinct admissible parameter
values produce the same distribution of $k_c$ when $N=4$.
\item[(iii)] If $\rho$ is unknown and $N \ge 5$, $(\theta, \eps_0, \eps_1, \rho)$
are identifiable for all parameter values outside a set of Lebesgue measure zero.
\end{enumerate}
\end{theorem}

\noindent The proof, given in Appendix~\ref{app:proof-bb}, reduces the
distribution of $k_c$ to the first $N$ moments of the acceptance propensity and
characterizes the solutions of the resulting polynomial system. Part (iii) is
established by an exact algebraic certificate.

Theorem~\ref{thm:bb} has three practical implications. First, estimating the
correlation raises the number of ratings required for generic identification
from three to five. At $N = 4$ the model has as many parameters as the count
distribution has free probabilities, and the resulting system can have two
admissible solutions. Second, in the limit $\rho = 1$ all ratings of an item
coincide, each item contributes a single Bernoulli observation, and the
distribution of $k_c$ depends on the parameters only through
$(1-\theta)\eps_0 + \theta(1-\eps_1)$. The parameters are then not identifiable
for any $N$, and the simulations of Section~\ref{sec:sims} indicate that
estimation becomes difficult as $\rho$ approaches one. Third, part (i) shows
that, for identification, the difficulty introduced by the correlation lies
entirely in estimating $\rho$ itself. The
simulation study of Section~\ref{sec:sims} is designed around these three
conditions.

\paragraph{Covariates and treatment effects.}
The prevalence may depend on item covariates $X_c$ and a binary intervention
$T_c$ through $\mathrm{logit}(\theta_c) = \alpha_\theta + X_c\beta + \tau T_c$,
which yields the treatment effect $\tau$ on the logit scale and, under
consistency, ignorability of treatment given $X_c$, and positivity, the average
treatment effect $\eta = E[D_c(T_c=1) - D_c(T_c=0)]$ on the probability scale.
The error rates are assumed not to depend on $T_c$. The error
rates may likewise depend on an observed attribute of item difficulty $H_c$
through logit-linear models for $u$ and $v$ in Equation~\eqref{eq:uv}.
Appendix~\ref{app:ext} states this model and shows in Theorem~\ref{thm:glm} that,
with $N \ge 5$ and full column rank of the design matrices, all regression
coefficients are identifiable for almost every parameter value. A covariate on the error rates shifts their mean across items but
does not describe the dispersion around that mean, so it complements rather than
replaces the correlation parameter. Appendix~\ref{app:difficulty} confirms this
empirically.

\paragraph{Semi-supervised estimation.}
Suppose the true state is revealed for a subset $\mathcal{L}$ of items, selected
independently of the true state and the ratings given the covariates and with
positive probability at every covariate value, while the
remaining items $\mathcal{U}$ are unlabeled. Labeled items contribute the joint
likelihood $\Pr(D_c \mid \theta_c)\Pr(k_c \mid D_c)$ and unlabeled items
contribute Equation~\eqref{eq:bb}. Because $D_c$ is observed for labeled items,
their contribution is not invariant to relabeling. Theorem~\ref{thm:semisup} in
Appendix~\ref{app:ext} shows that the semi-supervised model is identifiable for
$N \ge 2$ without the restriction $\eps_0+\eps_1<1$, so it accommodates an LLM
that is systematically wrong. Without labels, a model with $J < 0$ is not
identifiable, and the model restricted to $J>0$ targets the relabeled parameter
(Appendix~\ref{app:ext}). With $J = 0$ the prevalence is not identifiable. The
labeled block may be upweighted by a factor $W$. Setting $W = 1$ yields an
ordinary posterior, and Section~\ref{sec:anchors} shows empirically that
upweighting does not improve accuracy.

\subsection{Priors and estimation}
\label{sec:estimation}

We adopt a fully Bayesian approach. The logit of the prevalence receives a
$\mathcal{N}(0, 1.5^2)$ prior, $u$ and $v$ receive $\mathrm{Beta}(1,1)$ priors,
and $\rho$ receives a $\mathrm{Beta}(2,2)$ prior. In the semi-supervised model
$\eps_0$ and $\eps_1$ receive $\mathrm{Beta}(1,1)$ priors directly.
To avoid numerical underflow in the beta-binomial density when $\eps s$ is close
to zero, $u$ and $v$ are bounded to $(10^{-3}, 1-10^{-3})$ and $\rho$ to
$(0.01, 0.99)$. Models are fitted in Stan \citep{carpenter2017stan} through
cmdstanr with four chains of 2000 iterations, refitting once with 6000
iterations when the convergence criteria are not met. Convergence is assessed on
model parameters using rank-normalized $\widehat{R}$ and effective sample size
\citep{vehtari2021rank}. Of the 182 fits underlying Section~\ref{sec:empirical},
181 met the criteria, 14 of them after the longer run. One fit, of the
difficulty-covariate model for Qwen3-1.7B on VerifyBench under Arm~B, did not
converge ($\widehat{R}=1.45$) and is excluded. Appendix~\ref{app:fitting} gives
the diagnostics.

\section{Empirical study}
\label{sec:empirical}

\subsection{Data, models and prompt conditions}
\label{sec:setup}

\paragraph{Task.}
We use VerifyBench and VerifyBench-Hard \citep{yan2026verifybench}, released
under an MIT license. Each instance consists of a reasoning question $q$, a
reference answer $g$, a candidate response $r$, and a binary human label $y$
recording whether $r$ agrees with $g$. Labels were assigned by at least two
annotators with adjudication by a third. The judging task is to decide, given
$(q,g,r)$, whether the response agrees with the reference. The task cannot be
solved by string matching. The benchmark authors report that a standard
rule-based verifier reaches 66.95\% accuracy on the easier split, failing mainly
on answers that are algebraically equivalent to the reference but written in a
different form.

\paragraph{Items.}
We sample 150 items from VerifyBench and 100 from VerifyBench-Hard with a fixed
seed. VerifyBench pairs each question with one correct and one incorrect
response, so exactly half of its items carry a correct response. We take at most
one response per question, since two responses to the same question would not be
independent items. VerifyBench-Hard contains 29\% correct responses, and we
sample it by simple random sampling so that this imbalance is preserved, because
the prevalence is one of the quantities we estimate. The realized prevalence is
$0.493$ in the VerifyBench sample and $0.240$ in the VerifyBench-Hard sample.
Appendix~\ref{app:task} gives further detail.

\paragraph{Judges.}
We use seven open-weight instruction-tuned models from three families with 1B to
32B parameters (Table~\ref{tab:judges}), served with vLLM on a single A100 GPU at
each model's released precision, with checkpoints pinned to commit hashes. Each
item receives $N=10$ ratings at temperature $0.7$ and top-$p$ $0.9$. Where a
model provides a switch for extended reasoning we disable it. Two of the seven
models reason by default and provide no such switch. The selection rule, fixed
before data collection, was to take open-weight instruction-tuned models from
roughly 1B to 32B parameters across three widely deployed families, subject to
permissive licensing and single-GPU serving, and to report every model run.
Table~\ref{tab:judges} also lists the total error rate predicted for each model
before data collection from accuracy figures published by the benchmark authors.
All five models with a published figure fall within $0.061$ of the prediction.

\paragraph{Prompt conditions.}
Three prompt conditions are applied to the same items. Arm~A uses the benchmark
authors' published judge prompt verbatim, so that our error rates are comparable
with theirs. Arm~B adds one instruction eliciting a difficulty rating on a scale
from 1 to 5, which supplies an item-level covariate. Arm~C uses the authors'
published prompt without the reference answer, so the model must solve the
problem itself before judging the response. The decision remains binary and the
human label remains the target, so error rates under Arm~C are comparable with
those under Arm~A. The benchmark authors report an accuracy cost of 5 to 18
percentage points for withholding the reference. Arm~C therefore produces
models with high error rates without changing the model, and it corresponds to
applications in which no reference answer is available. Appendix~\ref{app:prompts}
describes the three prompts.

In total we collect 52{,}500 judgments, of which 94.2\% yielded a parseable
verdict. Recomputing the total error rate with unparsed ratings counted as
positive, counted as negative, or dropped changes it by at most $0.061$, and we
drop them. Items with fewer than three parseable ratings are excluded from the
fits, which retains 5{,}046 of the 5{,}250 combinations of item and setting, and
the remaining items enter the likelihood with their own $N_c$
(Appendix~\ref{app:task}).

\begin{table*}[t]
\centering\small
\caption{Judges, with the total error rate $\eps_0+\eps_1$ predicted before data
collection and measured on VerifyBench under Arm~A. Predictions are derived from
accuracy figures published by \citet{yan2026verifybench}. Because exactly half of
that split carries a correct response, $\eps_0+\eps_1 = 2(1-\text{accuracy})$, so
the published accuracy implies the predicted value without any fitting. Two
models postdate the benchmark and have no published figure.}
\label{tab:judges}
\begin{tabular}{llrrrr}
\toprule
Judge & Family & Parameters & Predicted & Measured & Difference \\
\midrule
gemma-3-1b-it        & Gemma   & 1B   & 0.947 & 0.908 & $-0.039$ \\
Qwen3-1.7B           & Qwen    & 1.7B & 0.378 & 0.439 & $+0.061$ \\
Qwen3-8B             & Qwen    & 8B   & 0.120 & 0.177 & $+0.057$ \\
Qwen3.5-9B           & Qwen    & 9B   & n/a   & 0.109 & n/a \\
gpt-oss-20b          & GPT-OSS & 20B  & 0.104 & 0.110 & $+0.006$ \\
gemma-4-26B-A4B-it   & Gemma   & 26B  & n/a   & 0.094 & n/a \\
Qwen3-32B            & Qwen    & 32B  & 0.084 & 0.132 & $+0.048$ \\
\bottomrule
\end{tabular}
\end{table*}

\subsection{Correlation among repeated ratings}
\label{sec:independence}

If the $N$ ratings of an item were independent, the variance of the acceptance
fraction $k_c/N$ among items with the same true state would be $p(1-p)/N$. The
ratio of the observed variance to this value is the design effect of
Equation~\eqref{eq:de}, and $\Neff = N/\DE$. We compute it separately among items
whose response is correct and items whose response is incorrect, so that the
difference between the two groups does not inflate it
(Appendix~\ref{app:de-method}).

\begin{table*}[t]
\centering\small
\caption{Design effect, implied intraclass correlation, and effective sample size
under Arm~A on VerifyBench with $N=10$ ratings per item. The right-hand columns
evaluate $\Neff = N/(1+(N-1)\rho)$ at other numbers of ratings using the measured
$\rho$, including the limit $1/\rho$ as $N \to \infty$.}
\label{tab:de}
\begin{tabular}{lrrrrrrr}
\toprule
& & & \multicolumn{5}{c}{$\Neff$ with $N$ ratings} \\
\cmidrule(l){4-8}
Judge & $\DE$ & $\rho$ & $N{=}3$ & $N{=}5$ & $N{=}10$ & $N{=}20$ & $N\to\infty$ \\
\midrule
gemma-3-1b-it       & 8.96 & 0.885 & 1.08 & 1.10 & 1.12 & 1.12 & 1.13 \\
Qwen3-1.7B          & 5.53 & 0.503 & 1.50 & 1.66 & 1.81 & 1.89 & 1.99 \\
Qwen3-8B            & 7.40 & 0.711 & 1.24 & 1.30 & 1.35 & 1.38 & 1.41 \\
Qwen3.5-9B          & 8.31 & 0.812 & 1.14 & 1.18 & 1.20 & 1.22 & 1.23 \\
gpt-oss-20b         & 8.24 & 0.804 & 1.15 & 1.19 & 1.21 & 1.23 & 1.24 \\
gemma-4-26B-A4B-it  & 8.90 & 0.878 & 1.09 & 1.11 & 1.12 & 1.13 & 1.14 \\
Qwen3-32B           & 7.46 & 0.717 & 1.23 & 1.29 & 1.34 & 1.37 & 1.39 \\
\bottomrule
\end{tabular}
\end{table*}

Across all 21 combinations of model and prompt condition on VerifyBench, the
intraclass correlation lies between $0.50$ and $0.93$ and $\Neff$ lies between
$1.07$ and $1.81$. Ten ratings are therefore worth between one and two
independent ratings. Treating them as ten independent observations understates
the standard error by a factor of about $\sqrt{\DE}$, which is $2.7$ at the median
design effect across all 42 settings and $2.8$ at the median on VerifyBench. The
correlation shows no apparent relation to model size, family, or whether the
model reasons by default. Under Arm~A the largest design effect belongs to a 1B
model and the second largest to a 26B model.

\paragraph{Effective sample size at other numbers of ratings.}
Because $\Neff = N/(1+(N-1)\rho)$ increases to the finite limit $1/\rho$, the
correlation bounds the information available from repeated querying for any
number of ratings. The right-hand columns of Table~\ref{tab:de} show the
magnitude of this bound. For the most correlated model, increasing the number
of ratings from ten to twenty raises $\Neff$ only from $1.116$ to $1.123$, and
the limit is $1.130$. For the least correlated model the same increase raises $\Neff$ from
$1.809$ to $1.894$, toward a limit of $1.99$. Across all 42 combinations of model, prompt
condition and split, the limit ranges from $1.04$ to $3.13$. These values are
extrapolations under the exchangeable model of Equation~\eqref{eq:bb} fitted at
$N=10$ rather than measurements at other values of $N$.

\paragraph{Self-reported difficulty.}
The design effects above are computed within groups of items with the same true
state, so they reflect variation in the error rate from item to item. The natural
candidate to explain this variation is the difficulty rating that the model
reports under Arm~B. That rating is weakly and positively associated with
realized error, with Spearman correlations between $0.18$ and $0.31$, but
conditioning on it removes between 0\% and 35\% of the excess variance, and less
than 15\% for four of the six models for which the comparison is available
(Appendix~\ref{app:difficulty}).

\subsection{Calibration of the binomial and beta-binomial likelihoods}
\label{sec:coverage}

We evaluate each fitted model against the error rates and prevalence computed
directly from the human labels, which the model does not use.

\begin{table*}[t]
\centering\small
\caption{Coverage of nominal 95\% credible intervals for the prevalence $\theta$
and the error rates $\eps_0$ and $\eps_1$, evaluated against the values computed
from human labels, and mean interval width for $\eps_0$. VB denotes VerifyBench
and VBH denotes VerifyBench-Hard. The two likelihoods share the same identifying
restriction and priors and differ only in their treatment of correlation within
items.}
\label{tab:coverage}
\begin{tabular}{llrrrrr}
\toprule
Likelihood & Split & Settings & Cover $\theta$ & Cover $\eps_0$ & Cover $\eps_1$ & Width $\eps_0$ \\
\midrule
Binomial      & VB  & 21 & 0.619 & 0.000 & 0.238 & 0.043 \\
Binomial      & VBH & 21 & 0.524 & 0.000 & 0.000 & 0.057 \\
Beta-binomial & VB  & 21 & 0.905 & 0.905 & 1.000 & 0.526 \\
Beta-binomial & VBH & 21 & 0.905 & 0.571 & 0.667 & 0.382 \\
\bottomrule
\end{tabular}
\end{table*}

As Table~\ref{tab:coverage} shows, the binomial interval for the false acceptance
rate contains the value computed from human labels in none of the 42 settings.
Its intervals are about twelve times narrower than those of the beta-binomial
model on VerifyBench and are largely insensitive to the correlation that
determines how much information the ratings carry.

This miscalibration extends to point estimation. The binomial model
underestimates the total error rate in all 42 settings, with a median
shortfall of $0.310$. The beta-binomial model underestimates it in 26 of 42, and
its median error across all 42 settings is $0.131$. Strongly correlated ratings produce counts
concentrated at $k_c=0$ and $k_c=N$. The binomial likelihood accounts for this
concentration by driving both error rates toward zero and attributing the
remaining variation to the prevalence. Section~\ref{sec:sims} reproduces this
behavior in simulation, where the true parameters are known.

On VerifyBench-Hard the beta-binomial interval covers the false acceptance rate
in $0.571$ of settings and the false rejection rate in $0.667$. The shortfall is
concentrated in the seven settings with $J \le 0$, all under Arm~C, where the
identifying condition fails and the model restricted to $J>0$ cannot represent
the true parameter. These seven settings account for every missed interval for
$\eps_1$ and for seven of the nine missed intervals for $\eps_0$. Among the 14
VerifyBench-Hard settings with $J>0$, coverage is $0.857$ for $\eps_0$ and $1.000$
for $\eps_1$. The remaining two misses on VerifyBench-Hard and the two misses on
VerifyBench all occur for gemma-3-1b-it, with $J$ between $0.07$ and $0.16$ and
$\rho$ between $0.78$ and $0.96$, where identification is weakest
(Appendix~\ref{app:vbh}). Across all 35 settings with $J>0$, the beta-binomial
interval covers $\eps_0$ in 31 and $\eps_1$ in all 35.

\subsection{Recovery and the acceptance gap}
\label{sec:separation}

Equation~\eqref{eq:sep} identifies $J = 1-(\eps_0+\eps_1)$ with the difference
between the acceptance rates on correct and incorrect responses.
Table~\ref{tab:sep} groups the 42 settings by this quantity.

\begin{table*}[t]
\centering\small
\caption{Recovery by acceptance gap $J$ under the beta-binomial model. AUC measures
how well the posterior probability $\Pr(D_c=1\mid k_c)$ ranks items against the
human labels. The two error columns are mean absolute deviations from the values
computed from human labels for the prevalence and the total error rate.}
\label{tab:sep}
\begin{tabular}{lrrrr}
\toprule
Acceptance gap $J$ & Settings & AUC & Prevalence error & Total error rate error \\
\midrule
Below $0.2$       & 14 & 0.484 & 0.335 & 0.620 \\
$0.2$ to $0.5$    &  5 & 0.701 & 0.120 & 0.161 \\
$0.5$ to $0.8$    & 13 & 0.851 & 0.046 & 0.105 \\
Above $0.8$       & 10 & 0.963 & 0.027 & 0.101 \\
\bottomrule
\end{tabular}
\end{table*}

The rank correlation between the acceptance gap and AUC is $0.957$ across the 42
settings. Below a gap of about $0.2$ the posterior separates the two classes poorly,
and where $J \le 0$ it reverses the ranking. The mean AUC in this group is
$0.484$.

This pattern bears two practical consequences. First, on the interpretation of
the identifying condition, treating $J>0$ as a threshold that is either met or
not met is incomplete, because a model that barely satisfies it behaves very
differently from one that satisfies it comfortably. The smallest model has a
total error rate of $0.908$ under Arm~A on VerifyBench, which satisfies the
condition with $J = 0.092$, yet recovers the latent state with an AUC of only $0.559$. Among the
35 settings in which the condition holds, the lowest AUC is $0.544$, at
$J = 0.083$. The seven settings that rank items worse than chance are exactly
those with $J \le 0$, where the condition fails. The second is
practical. The acceptance gap can be estimated from a small labeled sample, so a
practitioner can assess in advance whether an unsupervised fit is likely to be
informative.

\subsection{The value of gold labels}
\label{sec:anchors}

We examine the value of gold labels in two ways. First, in each of the 42
settings we fit the semi-supervised beta-binomial model with the true state
revealed for 10\% of items, drawn stratified on the true state
(Appendix~\ref{app:strat}), and compare it with the unsupervised beta-binomial
fit to the same setting. Second, in nine settings chosen to span the range of
acceptance gaps, three each with $J$ near $0.16$, $0.62$ and $0.91$, we vary the
labeled fraction over $\{5,10,20,35,50\}\%$ and the weight on labeled items
over $W \in \{1, 3, W_{\text{eq}}\}$, where
$W_{\text{eq}}=\max(1, \lfloor|\mathcal{U}|/|\mathcal{L}|\rfloor)$ gives the
labeled and unlabeled blocks approximately equal total weight, with five draws of the labeled subset at
each fraction. In both analyses the semi-supervised model does not impose
$\eps_0+\eps_1<1$, and ranking accuracy is evaluated on unlabeled items only.
The acceptance gap $J$ used to group settings is computed from the human labels.

Stratifying on the true state with proportional allocation makes the prevalence
among labeled items close to the sample prevalence. This uses information that a
random labeled sample of the same size would not provide, and it departs from
the selection condition of Theorem~\ref{thm:semisup}. The reductions in
prevalence error reported below therefore overstate what a simple random sample
of labels would achieve, particularly at the smallest labeled fractions. The
simulation in Appendix~\ref{app:sim-semisup}, which draws labels by simple random
sampling, shows the same qualitative pattern with 10\% of items labeled.

\begin{table*}[t]
\centering\small
\caption{Unsupervised and semi-supervised beta-binomial fits to the same 42
settings, grouped by whether the identifying condition $J>0$ holds. The
semi-supervised fits reveal 10\% of items, drawn stratified on the true state,
with $W = W_{\text{eq}}$. Prevalence error is the mean absolute deviation from the
prevalence computed from human labels, and AUC is computed on unlabeled items.}
\label{tab:twobytwo}
\begin{tabular}{llrrrr}
\toprule
Identification & Supervision & Settings & AUC & Prevalence error & Cover $\theta$ \\
\midrule
$J > 0$     & Unsupervised      & 35 & 0.811 & 0.122 & 0.886 \\
$J > 0$     & Semi-supervised   & 35 & 0.753 & 0.027 & 0.943 \\
$J \le 0$   & Unsupervised      &  7 & 0.368 & 0.273 & 1.000 \\
$J \le 0$   & Semi-supervised   &  7 & 0.406 & 0.040 & 0.857 \\
\bottomrule
\end{tabular}
\end{table*}

Table~\ref{tab:twobytwo} shows that labeling 10\% of items reduces the error in
the estimated prevalence by a factor of between four and seven, from $0.122$ to
$0.027$ where $J>0$ and from $0.273$ to $0.040$ where $J \le 0$. Ranking accuracy
does not improve, and where $J>0$ the AUC falls from $0.811$ to $0.753$ for the
reason discussed below. These fits weight labeled items by $W_{\text{eq}}$. The
experiment below shows that accuracy does not depend on $W$ and that intervals
are wider with $W=1$.

\begin{table*}[t]
\centering\small
\caption{Effect of the labeled fraction on the error in the estimated prevalence
and on ranking accuracy, with $W=1$, by acceptance gap. Each entry averages three
settings and five draws of the labeled subset. Prevalence error is the mean
absolute deviation from the prevalence computed from human labels, and AUC is
computed on unlabeled items.}
\label{tab:anchors}
\begin{tabular}{lrrrrrr}
\toprule
& \multicolumn{2}{c}{$J \approx 0.16$} & \multicolumn{2}{c}{$J \approx 0.62$} & \multicolumn{2}{c}{$J \approx 0.91$} \\
\cmidrule(lr){2-3} \cmidrule(lr){4-5} \cmidrule(lr){6-7}
Labeled fraction & Error & AUC & Error & AUC & Error & AUC \\
\midrule
0\%  & 0.286 & 0.650 & 0.025 & 0.823 & 0.019 & 0.967 \\
5\%  & 0.030 & 0.517 & 0.040 & 0.820 & 0.014 & 0.971 \\
10\% & 0.030 & 0.591 & 0.029 & 0.827 & 0.025 & 0.966 \\
20\% & 0.035 & 0.594 & 0.024 & 0.807 & 0.017 & 0.967 \\
35\% & 0.009 & 0.629 & 0.030 & 0.792 & 0.012 & 0.967 \\
50\% & 0.010 & 0.648 & 0.021 & 0.791 & 0.008 & 0.960 \\
\bottomrule
\end{tabular}
\end{table*}

Table~\ref{tab:anchors} shows the same result as a function of the labeled
fraction. Where the acceptance gap is about $0.16$, labeling 5\% of items, that
is five to seven items drawn stratified on the true state, reduces the error in
the estimated prevalence from $0.286$ to $0.030$, which is 93\% of the reduction
obtained when half of the items are labeled. Where the gap is about $0.62$ or $0.91$, the unsupervised
estimate is already within $0.025$ of the reference value and labeling changes
it little.

Labeling does not improve ranking accuracy. Where the gap is about $0.16$ the
mean AUC falls from $0.650$ without labels to $0.517$ with 5\% of items
labeled, and returns to $0.648$ when half of the items are labeled. The fall
occurs because the semi-supervised model does not impose $J>0$, and with few
labels and weakly separated classes it sometimes estimates the direction of
separation incorrectly. Six of the fifteen fits with 5\% of items labeled rank
items in reverse order. More generally, when the two acceptance rates are close,
the count $k_c$ of an individual item is nearly as likely under either true
state, and labels on other items cannot supply the information about that item
which its own ratings do not contain. The prevalence, in contrast, is
constrained directly by the labeled subset. A labeled subset is
therefore most valuable where the unsupervised model is least reliable, provided
the target is a population quantity. When the target is the classification of
individual items, labeling does not substitute for a model that separates the
two classes more clearly.

Across the experiment, the mean absolute error in the estimated prevalence is
$0.022$, $0.020$ and $0.022$ for $W=1$, $W=3$ and $W=W_{\text{eq}}$
respectively, while the mean interval width decreases from $0.292$ to $0.190$ to
$0.170$. Setting $W>1$ raises the labeled likelihood to a power greater than
one, which does not correspond to a valid Bayesian update and produces intervals
narrower than the evidence supports without improving accuracy. We therefore
recommend $W=1$ (Appendix~\ref{app:weight}).

\section{Simulation study}
\label{sec:sims}

The empirical study measures recovery against human labels, at a single number of
ratings, and at the correlations the models happen to exhibit. The simulation
study examines the estimators where the true parameters are known, and its design
follows the identification conditions of Theorem~\ref{thm:bb}.

\subsection{Design}
\label{sec:sim-design}

Data are generated from Equation~\eqref{eq:bb} with $\theta = 0.3$,
$\eps_0 = 0.15$ and $\eps_1 = 0.10$, so that $J = 0.75$, and with $C = 250$ items,
the total size of the benchmark sample. The number of ratings takes the values
$N \in \{3, 4, 5, 10, 20\}$. With the correlation unknown, $N = 3$ fails the
necessary condition of Theorem~\ref{thm:bb}(ii). $N = 4$ satisfies it but is not
sufficient, and for this parameter the system at $N = 4$ has a second admissible
solution when $\rho \in \{0.75, 0.90\}$ (Appendix~\ref{app:proof-bb}). Values
$N \ge 5$ satisfy Theorem~\ref{thm:bb}(iii). The intraclass correlation takes the
values $\rho \in \{0, 0.25, 0.50, 0.75, 0.90\}$, which span the range measured in
Section~\ref{sec:independence} and approach the boundary $\rho = 1$ at which
identification fails. Each setting is replicated 250 times.

We compare four procedures. The binomial and beta-binomial latent class models
are fitted by maximum likelihood under the parameterization of
Equation~\eqref{eq:uv}, with Wald intervals formed on the unconstrained scale. We
use maximum likelihood rather than the Bayesian estimator of
Section~\ref{sec:estimation} so that many replicates can be run. Majority vote
assigns each item its majority label, estimates the prevalence as the proportion
of positive labels, and estimates the error rates against the majority label.
Single-round classification uses the first rating of each item as its label. For
the last two procedures we report the Wald interval for a proportion that a
practitioner would compute by treating the assigned labels as correct.
Appendix~\ref{app:sim-main} gives further detail. The simulation concerns the
measurement parameters $(\theta, \eps_0, \eps_1, \rho)$ and the conditions of
Theorem~\ref{thm:bb}. It does not include covariates or a treatment, and the
regression and causal extensions depend on Theorem~\ref{thm:glm} without numerical
evaluation in this paper.

\subsection{Results}
\label{sec:sim-results}

\begin{table*}[t]
\centering\small
\setlength{\tabcolsep}{4.5pt}
\caption{Bias of the estimated total error rate $\eps_0+\eps_1$, whose true value is $0.25$, and coverage of nominal 95\% intervals, by intraclass correlation $\rho$ and number of ratings $N$. $C = 250$ items and 250 replicates per setting. $^{\dagger}$At $N=3$ the observed information of the beta-binomial model is singular in about half of the replicates, and coverage is computed on the remainder.}
\label{tab:sim-total}
\begin{tabular}{llrrrrrrrr}
\toprule
 & & \multicolumn{2}{c}{$N=3$} & \multicolumn{2}{c}{$N=4$} & \multicolumn{2}{c}{$N=5$} & \multicolumn{2}{c}{$N=10$} \\
\cmidrule(lr){3-4} \cmidrule(lr){5-6} \cmidrule(lr){7-8} \cmidrule(lr){9-10}
$\rho$ & Likelihood & Bias & Cover & Bias & Cover & Bias & Cover & Bias & Cover \\
\midrule
0.00 & Binomial & $-0.003$ & 0.95 & $-0.002$ & 0.95 & $-0.001$ & 0.94 & $-0.002$ & 0.96 \\
 & Beta-binomial & $+0.008$ & 1.00$^{\dagger}$ & $+0.001$ & 0.97 & $0.000$ & 0.96 & $-0.002$ & 0.96 \\
\addlinespace[2pt]
0.25 & Binomial & $-0.054$ & 0.51 & $-0.046$ & 0.52 & $-0.039$ & 0.48 & $-0.021$ & 0.59 \\
 & Beta-binomial & $+0.008$ & 1.00$^{\dagger}$ & $-0.008$ & 0.88 & $-0.004$ & 0.93 & $+0.004$ & 0.95 \\
\addlinespace[2pt]
0.50 & Binomial & $-0.120$ & 0.00 & $-0.104$ & 0.01 & $-0.104$ & 0.02 & $-0.087$ & 0.01 \\
 & Beta-binomial & $-0.046$ & 1.00$^{\dagger}$ & $-0.036$ & 0.75 & $-0.032$ & 0.71 & $-0.016$ & 0.88 \\
\addlinespace[2pt]
0.75 & Binomial & $-0.185$ & 0.00 & $-0.177$ & 0.00 & $-0.177$ & 0.00 & $-0.168$ & 0.00 \\
 & Beta-binomial & $-0.125$ & 0.98$^{\dagger}$ & $-0.130$ & 0.53 & $-0.115$ & 0.58 & $-0.100$ & 0.56 \\
\addlinespace[2pt]
0.90 & Binomial & $-0.225$ & 0.00 & $-0.219$ & 0.00 & $-0.221$ & 0.00 & $-0.217$ & 0.00 \\
 & Beta-binomial & $-0.199$ & 0.88$^{\dagger}$ & $-0.201$ & 0.33 & $-0.199$ & 0.37 & $-0.187$ & 0.36 \\
\bottomrule
\end{tabular}
\end{table*}

Table~\ref{tab:sim-total} reports the estimated total error rate. When $\rho = 0$
both likelihoods are unbiased and their intervals attain nominal coverage. As
$\rho$ increases, the binomial model underestimates the total error rate, by
between $0.08$ and $0.23$ when $\rho \ge 0.50$, and the coverage of its interval
falls to $0.02$ or less. Increasing the number of ratings does not reduce this
bias. At $\rho = 0.90$ the bias is $-0.225$ with three ratings and $-0.216$ with
twenty. The binomial likelihood is therefore inconsistent under correlated
ratings, which reproduces the result of Section~\ref{sec:coverage} in a setting
where the truth is known.

The behavior of the beta-binomial model follows Theorem~\ref{thm:bb}. With
$N = 3$ its observed information is singular in about half of the
replicates, and the intervals that can be formed are wide enough to cover almost
always. With $N = 4$ the model has small bias when $\rho \le 0.25$ but is biased
by $-0.130$ and $-0.201$ at $\rho = 0.75$ and $\rho = 0.90$, which are the two
settings in which the parameter is not identified at $N = 4$. With $N \ge 5$ the
model is essentially unbiased with nominal coverage when $\rho \le 0.25$, and at
$\rho = 0.50$ its bias is between $-0.016$ and $-0.032$ with coverage between
$0.71$ and $0.88$. At $\rho = 0.75$ and $\rho = 0.90$ a downward bias of about
$0.10$ and $0.19$ remains for every $N \ge 5$, nearly as large as at $N = 4$, so
the table does not separate the exact failure at $N = 4$ from weak
identification at $N \ge 5$. Both are consistent with the loss of identification
at $\rho = 1$. For this parameter the second solutions at $N = 4$ have lower
error rates and a lower correlation than the truth
(Appendix~\ref{app:proof-bb}), which matches the direction of the observed
bias. The beta-binomial model thus removes most of the bias at moderate
correlation and a smaller part of it at high correlation. Results at $N = 20$
are close to those at $N = 10$ (Appendix~\ref{app:sim-main}).

\begin{table*}[t]
\centering\small
\caption{Comparison of procedures with $N = 10$ ratings per item, by intraclass correlation $\rho$. Prevalence coverage for majority vote and single round uses a Wald interval that treats the assigned labels as correct. Majority vote estimates the error rates against the majority label, and a single round does not estimate them. Accuracy is the proportion of items classified correctly.}
\label{tab:sim-compare}
\begin{tabular}{lrrrrr}
\toprule
 & \multicolumn{5}{c}{$\rho$} \\
\cmidrule(l){2-6}
Procedure & 0.00 & 0.25 & 0.50 & 0.75 & 0.90 \\
\midrule
\multicolumn{6}{l}{\emph{Prevalence bias}} \\
Binomial & $-0.006$ & $+0.041$ & $+0.073$ & $+0.076$ & $+0.084$ \\
Beta-binomial & $-0.006$ & $-0.012$ & $-0.016$ & $+0.053$ & $+0.082$ \\
Majority vote & $-0.001$ & $+0.044$ & $+0.066$ & $+0.072$ & $+0.082$ \\
Single round & $+0.072$ & $+0.073$ & $+0.076$ & $+0.075$ & $+0.082$ \\
\addlinespace[3pt]
\multicolumn{6}{l}{\emph{Prevalence coverage}} \\
Binomial & 0.96 & 0.73 & 0.35 & 0.29 & 0.20 \\
Beta-binomial & 0.96 & 0.92 & 0.75 & 0.69 & 0.44 \\
Majority vote & 0.95 & 0.67 & 0.40 & 0.34 & 0.21 \\
Single round & 0.34 & 0.31 & 0.28 & 0.31 & 0.20 \\
\addlinespace[3pt]
\multicolumn{6}{l}{\emph{Total error rate bias}} \\
Binomial & $-0.002$ & $-0.021$ & $-0.087$ & $-0.168$ & $-0.217$ \\
Beta-binomial & $-0.002$ & $+0.004$ & $-0.016$ & $-0.100$ & $-0.187$ \\
Majority vote & $+0.001$ & $-0.019$ & $-0.090$ & $-0.171$ & $-0.218$ \\
\addlinespace[3pt]
\multicolumn{6}{l}{\emph{Classification accuracy}} \\
Binomial & 0.998 & 0.929 & 0.881 & 0.865 & 0.864 \\
Beta-binomial & 0.998 & 0.938 & 0.897 & 0.868 & 0.863 \\
Majority vote & 0.996 & 0.927 & 0.885 & 0.867 & 0.865 \\
Single round & 0.863 & 0.865 & 0.864 & 0.862 & 0.865 \\
\bottomrule
\end{tabular}
\end{table*}

Table~\ref{tab:sim-compare} compares the four procedures with $N = 10$. A single
round overestimates the prevalence by about $0.075$ at every $\rho$, which equals
$(1-\theta)\eps_0 + \theta(1-\eps_1) - \theta$, and its interval covers the truth
in at most 34\% of replicates. Majority vote is accurate when $\rho = 0$, but
its advantage over a single round declines as $\rho$ increases. At $\rho = 0.90$
the two procedures have the same prevalence bias of $0.082$ and nearly the same
classification accuracy, consistent with an effective sample size close to one.
The error rates that majority vote implies are as biased as those of the binomial
model, and the prevalence estimated by the binomial model is as biased as that of
majority vote. At $\rho = 0.50$ the beta-binomial model estimates the prevalence
with a bias of $-0.016$ and coverage of $0.75$, against $0.35$ for the binomial
model, $0.40$ for majority vote and $0.28$ for a single round, and it classifies
items slightly more accurately. At $\rho = 0.90$ all four procedures estimate the
prevalence with similar bias. The beta-binomial interval covers the truth in
$0.44$ of replicates against about $0.20$ for the others.

The coverage figures in this section use Wald intervals from the observed
information rather than the posterior intervals used in
Section~\ref{sec:empirical}, and the two need not agree. On VerifyBench the
posterior intervals attained coverage of $0.905$ for both the prevalence and the
false acceptance rate.

\subsection{Semi-supervised estimation and prediction-powered inference}
\label{sec:sim-semisup}

Appendix~\ref{app:sim-semisup} reports a second simulation study of the
semi-supervised model with a 10\% labeled subset, compared with
prediction-powered inference (PPI), estimation from the labeled items alone, the
unsupervised model, and majority vote, at acceptance gaps $J = 0.75$, $0.20$ and
$-0.30$. The semi-supervised model and PPI are both
approximately unbiased with nominal coverage in all three settings, and the
semi-supervised model has the smaller root mean squared error. The unsupervised
model overestimates the prevalence by $0.21$ when $J = 0.20$, where the classes
are weakly separated, and by $0.30$ when $J = -0.30$. In the latter setting the
model is not identified without labels, and as the remark following
Theorem~\ref{thm:semisup} anticipates, its estimate moves toward the relabeled
prevalence $1-\theta = 0.70$, which differs from the truth by $0.40$. As in
Section~\ref{sec:anchors}, labels do not improve ranking accuracy when $J > 0$.
When $J < 0$ they restore the direction of the ranking, raising the AUC from
$0.27$ to $0.72$.

\section{Discussion}
\label{sec:discussion}

This paper examined the assumption, common to repeated querying and to standard
latent class models, that repeated LLM ratings of the same item are
conditionally independent. On a public benchmark with human labels, the
assumption fails for every model and prompt condition we examined, and on the
balanced split ten ratings carry the information of between one and two
independent ratings. A
binomial latent class model fitted under this correlation reports intervals that
are too narrow and error rates that are too low. The proposed beta-binomial
latent class model accounts for the correlation with a single parameter, is
identifiable under the conditions of Theorem~\ref{thm:bb}, and on the balanced split of
the benchmark raises interval coverage close to the nominal level. Recovery of the latent state is governed by the
acceptance gap $J$, and a labeled subset improves population estimates where $J$
is small without improving the classification of individual items.

These results suggest the following practice. First, estimate the acceptance gap
from a small labeled sample before relying on an unsupervised fit. Below about
$0.5$ the classification of individual items degrades markedly, below about
$0.2$ it is unreliable, and where $J \le 0$ the unsupervised model reverses the
ranking. Second, use the beta-binomial likelihood unless
the design effect of Equation~\eqref{eq:de}, computed within groups of items with
a known label, is close to one. Third, collect at least five ratings per item
when the correlation is estimated, while recognizing that ratings beyond ten add
little information. Fourth, choose the estimator according to the estimand. When
the target is a population rate and a probability sample of gold labels is
available, prediction-powered inference and design-based supervised learning
reach it under weaker assumptions. The latent class model is appropriate when
posterior probabilities for individual items, error rates, or treatment effects
are also required. Finally, when the acceptance gap is small and the target is a
population rate, additional gold labels are more valuable than additional
ratings. When the target is the classification of individual items, neither
helps, and the model itself must be improved.

\subsection{Limitations}
\label{sec:limitations}

The empirical study covers 250 items from a single benchmark family, one task
type, seven open-weight models, one sampling configuration, and one number of
ratings per item. We do not claim that the correlations we measure transfer to
other tasks, to proprietary models, or to other sampling configurations. The
study establishes that strong correlation within items occurs, that it occurs in
every setting we examined, and that it is sufficient to invalidate the binomial
likelihood. The numerical values reported here should be read as descriptive of
this setting.

We treat the human labels as ground truth. While inspecting items on which models
disagreed most, we found cases whose published label was difficult to
adjudicate, and the benchmark authors' own error analysis discusses a similar
case. We did not audit the labels systematically. Where the true label is itself
uncertain, part of what we attribute to model error would be better attributed to
annotation error, which is the concern raised by \citet{guerdan2025validating}.

Twenty-six of the 250 items have responses generated by gemma-3-1b-it, which is
also one of the seven judges. No other judge generated any response in the
sample. We have not compared error rates on these items with those on the
remaining items. VerifyBench-Hard contains only about 24 items with a
correct response, so error rates conditional on a correct response are imprecise
on that split. Finally, in the labeling experiment of Section~\ref{sec:anchors}
labeled items were drawn stratified on the true state, which overstates the
value of small labeled subsets relative to simple random sampling.

\subsection{Future work}
\label{sec:future}

These findings suggest several empirical and methodological extensions. Additional benchmarks
from other domains, with different annotation protocols, would establish whether
the correlation we measure is a general property of LLM judgments. Collecting
ratings at several values of $N$ would test the exchangeability assumption of
Equation~\eqref{eq:bb} directly, which is not possible with a single $N$. Repeating the labeling experiment with simple random
sampling of labeled items would measure the value of small labeled subsets
without the advantage conferred by stratification.
A simulation study of the treatment effect $\tau$ under correlated ratings would
evaluate the causal extension numerically. Methodologically, the model could be
extended to multinomial and ordinal ratings, where the beta-binomial component would be replaced by a
Dirichlet-multinomial component, and to joint modeling of annotator error, in
which the human labels are treated as a second imperfect measurement rather than
as the truth.

\section*{Acknowledgments}

We are grateful to Eric Culbertson, Grace Deng, Yuhe Gao, Shinpei
Nakamura-Sakai, Bahman Rabii, Jan Vlachy, and Ka Wong for helpful comments.

\clearpage
\appendix


\section{Proofs and model extensions}
\label{app:proofs}

\subsection{Proof of Theorem~\ref{thm:binom}}
\label{app:proof-binom}

Let $k_c = \sum_{i=1}^{N_c} R_{c,i}$. Under conditional independence the
marginal distribution of $k_c$ is the binomial mixture of
Equation~\eqref{eq:binom}. By the identifiability of finite binomial mixtures
\citep{blischke1964estimating,teicher1963identifiability}, a mixture of $K$
binomial distributions with distinct success probabilities is identifiable up
to relabeling of the components if and only if $N_c \ge 2K-1$. For a binary
latent state $K=2$, which requires $N_c \ge 3$.

The likelihood is invariant under the relabeling map
$\mathcal{T}(\theta,\eps_0,\eps_1) = (1-\theta,\,1-\eps_1,\,1-\eps_0)$, since exchanging
which component is called positive exchanges the roles of the two error rates
and complements the mixing weight. Under $\mathcal{T}$ the total error rate becomes
$(1-\eps_1)+(1-\eps_0) = 2-(\eps_0+\eps_1)$. If $\eps_0+\eps_1<1$, the image
therefore satisfies $2-(\eps_0+\eps_1)>1$ and lies outside the restricted
parameter space. Exactly one member of each relabeled pair is admissible, which
removes the remaining ambiguity \citep{stephens2000dealing}. Because the two
components have acceptance rates $\eps_0$ and $1-\eps_1$, the restriction states
that $J = (1-\eps_1)-\eps_0 > 0$. \hfill$\square$

\subsection{Proof of Theorem~\ref{thm:bb}}
\label{app:proof-bb}

Under Equation~\eqref{eq:bb}, each item has an acceptance propensity $\pi_c$ and
$k_c \mid \pi_c \sim \Bin(N, \pi_c)$. Conditional on $D_c = d$, $\pi_c$ follows a
$\mathrm{Beta}(\mu_d s, (1-\mu_d)s)$ distribution with $\mu_0 = \eps_0$ and
$\mu_1 = 1-\eps_1$, where the case $\rho = 0$ is understood as the limit
$s \to \infty$, in which $\pi_c = \mu_d$. To avoid confusion with the item index
$c$, write $\kappa = \rho/(1-\rho) = 1/s$, which maps $[0,1)$ one to one and
smoothly onto $[0,\infty)$, so that a set of parameter values has Lebesgue
measure zero in terms of $\rho$ if and only if it does in terms of $\kappa$. Let
$m_r = E[\pi_c^r]$ for $r = 0, 1, \dots, N$, with $m_0 = 1$. Let $G$ denote the
distribution placing mass $1-\theta$ at $\mu_0$ and mass $\theta$ at $\mu_1$, and
write $g_r = E_G[\mu^r]$.

\begin{lemma}
\label{lem:moments}
For fixed $N$, the distribution of $k_c$ determines $(m_1,\dots,m_N)$ and is
determined by it.
\end{lemma}

\begin{proof}
For $k = 0, \dots, N$,
\begin{align*}
\Pr(k_c = k) &= \binom{N}{k} E[\pi^k(1-\pi)^{N-k}]\\
&= \binom{N}{k}\sum_{j=0}^{N-k}(-1)^j\binom{N-k}{j} m_{k+j},
\end{align*}
and conversely $E[k_c(k_c-1)\cdots(k_c-r+1)] = N(N-1)\cdots(N-r+1)\, m_r$ for
$r \le N$.
\end{proof}

\begin{lemma}
\label{lem:rising}
Let $P_r(\mu; \kappa) = \prod_{j=0}^{r-1}(\mu + j\kappa)$ and
$Q_r(\kappa) = \prod_{j=0}^{r-1}(1 + j\kappa)$. Then
$Q_r(\kappa)\, m_r = E_G[P_r(\mu; \kappa)]$ for every $r \ge 1$. Consequently, for
any value $\kappa'$, the quantities
\[
g_r(\kappa') = Q_r(\kappa')\, m_r - \sum_{i=0}^{r-1} p_{r,i}(\kappa')\, g_i(\kappa'),
\]
with $g_0(\kappa') = 1$,
where $p_{r,i}(\kappa')$ is the coefficient of $\mu^i$ in $P_r(\mu; \kappa')$, are
polynomials in $\kappa'$, and $g_r(\kappa) = g_r$ at the true value $\kappa$.
\end{lemma}

\begin{proof}
The $r$-th moment of a $\mathrm{Beta}(\mu s, (1-\mu)s)$ distribution is
\[
\prod_{j=0}^{r-1} \frac{\mu s + j}{s + j} \;=\; \frac{P_r(\mu; \kappa)}{Q_r(\kappa)}.
\]
Averaging over $G$ gives the first statement. Because $P_r$ is monic of degree $r$ in $\mu$, the
relation $Q_r(\kappa) m_r = \sum_{i\le r} p_{r,i}(\kappa) g_i$ can be solved
recursively for $g_r$ with no division, which gives the second.
\end{proof}

\begin{lemma}
\label{lem:twopoint}
Let $g_0,\dots,g_{N}$ be the moments of a distribution placing positive mass on
exactly two points $x_1 < x_2$. Then (a) the $3 \times (N-1)$ Hankel matrix with
entries $g_{i+j}$, $i = 0,1,2$, $j = 0,\dots,N-2$, has rank at most two, and (b)
$g_1, g_2, g_3$ determine $x_1$, $x_2$ and the weights.
\end{lemma}

\begin{proof}
For (a), each row of the Hankel matrix is $w_1 x_1^i (1, x_1, x_1^2,\dots) +
w_2 x_2^i(1, x_2, x_2^2, \dots)$, so the rows lie in a space of dimension two. For
(b), $x_1$ and $x_2$ are the roots of $x^2 - e_1 x + e_2$, where $(e_1, e_2)$
solve $g_2 = e_1 g_1 - e_2$ and $g_3 = e_1 g_2 - e_2 g_1$. The determinant of this
linear system is $g_2 - g_1^2$, the variance of the distribution, which is
positive. The weight on $x_2$ is $(g_1 - x_1)/(x_2 - x_1)$.
\end{proof}

\paragraph{Part (i).}
If $\rho$ is known then $\kappa$ is known, and Lemmas~\ref{lem:moments}
and~\ref{lem:rising} give $g_1, g_2, g_3$ from the distribution of $k_c$ when
$N \ge 3$. Lemma~\ref{lem:twopoint}(b) then determines the two support points
$\{\mu_0, \mu_1\}$ and their weights. Since $J > 0$ is equivalent to
$\mu_0 < \mu_1$, the smaller support point is $\eps_0$, the larger is
$1-\eps_1$, and the weight on the larger is $\theta$.

\paragraph{Part (ii).}
For $N \le 3$ the distribution of $k_c$ has at most three free probabilities,
while the parameter space has dimension four. A continuous injective map from an
open subset of $\mathbb{R}^4$ into $\mathbb{R}^3$ does not exist, by invariance of
domain, so $N \ge 4$ is necessary.

To show that $N = 4$ is not sufficient, take $\theta = 1/2$, $\eps_0 = 3/20$,
$\eps_1 = 1/10$ and $\rho = 1/2$, so that $\kappa = 1$. Eliminating the support
points from the moment equations for $m_1,\dots,m_4$ yields a polynomial
condition on the correlation parameter, whose roots other than the true value
$\kappa = 1$ are the roots of $686542\, \kappa'^2 - 3824088\, \kappa' + 5264471$.
The root $\kappa' = 2.48784$ corresponds to the admissible parameter
\begin{align*}
\theta' &= 0.25132, & \eps_0' &= 0.38341,\\
\eps_1' &= 0.05320, & \rho' &= 0.71329,
\end{align*}
with $J' = 0.5634 > 0$. Computation with more than 150 significant digits
confirms that this parameter produces the same moments $m_1,\dots,m_4$ as the
true parameter, and hence by Lemma~\ref{lem:moments} the same distribution of
$k_c$ at $N = 4$. The two distributions differ at $N = 5$. The other root,
$\kappa' = 3.08223$, corresponds to a negative false rejection rate and is not
admissible. Because $\kappa' = 2.48784$ is a simple root of a polynomial whose
coefficients depend continuously on the true parameter, it persists under small
perturbations of the true parameter, and the admissibility conditions are strict
inequalities that also persist. The set of parameter values with a second
admissible solution at $N = 4$ therefore contains an open neighborhood of this
point and has positive Lebesgue measure.

\paragraph{Part (iii).}
Suppose an admissible parameter $(\theta', \eps_0', \eps_1', \rho')$ produces the
same distribution of $k_c$ as the true parameter with $N \ge 5$, and let
$\kappa' = \rho'/(1-\rho')$. By Lemmas~\ref{lem:moments} and~\ref{lem:rising}, the
sequence $g_0(\kappa'), \dots, g_5(\kappa')$ computed from the common moments is
the moment sequence of the two-point distribution defined by the alternative
parameter, so by Lemma~\ref{lem:twopoint}(a) the $3 \times 4$ Hankel matrix built
from it has rank at most two. Let $M_1(\kappa')$ and $M_2(\kappa')$ denote its
$3\times 3$ minors formed from columns one to three and two to four. The moments
$m_r$ are rational functions of $(\theta, \eps_0, \eps_1, \kappa)$ whose
denominators $Q_r(\kappa)$ are positive. After multiplication by these positive
factors, $M_1$ and $M_2$ are polynomials in $\kappa'$ and in
$(\theta, \eps_0, \eps_1, \kappa)$, of degree at most four and six in $\kappa'$.

Both minors vanish at $\kappa' = \kappa$, because $g(\kappa)$ is the moment
sequence of $G$. Both also vanish at $\kappa' = -1$, a value that corresponds to
no $\rho \in [0,1)$. At that value $Q_r(-1) = 0$ for $r \ge 2$ and
$P_r(\mu; -1) = \mu(\mu-1)\cdots(\mu-r+1)$, whose coefficients sum to zero and
whose constant term is zero, so the recursion of Lemma~\ref{lem:rising} gives
$g_r(-1) = m_1$ for every $r \ge 1$. This is the moment sequence of a Bernoulli
distribution, whose Hankel matrix has rank at most two. Because
$(\kappa' - \kappa)(\kappa' + 1)$ is monic in $\kappa'$, it divides both minors
with a quotient whose coefficients are again polynomials in the parameters,
$M_j(\kappa') = (\kappa' - \kappa)(\kappa' + 1)\, S_j(\kappa')$.

Let $\mathrm{Res}$ denote the resultant of $S_1$ and $S_2$ with respect to
$\kappa'$, taken with their formal degrees two and four, which is a polynomial in
$(\theta, \eps_0, \eps_1, \kappa)$. At the rational parameter $\theta = 1/2$,
$\eps_0 = 3/20$, $\eps_1 = 1/10$, $\rho = 1/2$, exact computation gives $S_1$ and
$S_2$ of degrees exactly two and four and a nonzero resultant, so the polynomial
$\mathrm{Res}$ is not identically zero. The same holds at
$(\theta, \eps_0, \eps_1, \rho) = (3/10, 1/10, 1/5, 4/5)$ and at
$(7/10, 2/5, 1/5, 1/4)$. The zero set $\mathcal{Z}$ of a polynomial that is not
identically zero has Lebesgue measure zero. For any true parameter outside
$\mathcal{Z}$, $S_1$ and $S_2$ have no common root in $\mathbb{C}$, since a
common root would force the resultant to vanish whether or not a leading
coefficient vanishes. The only common roots of $M_1$ and $M_2$ are therefore
$\kappa$ and $-1$. An admissible alternative has $\kappa' \ge 0$ and therefore
$\kappa' = \kappa$, and part (i) then gives
$(\theta', \eps_0', \eps_1') = (\theta, \eps_0, \eps_1)$. For $N > 5$ the same two
minors are available, so the conclusion holds for every $N \ge 5$.
\hfill$\square$

The computations are reproduced by scripts in the code repository.
\texttt{exact\_n4.py} computes the counterexample of part (ii),
\texttt{exact\_cert3.py} computes the resultants of part (iii), and
\texttt{scan\_n4.py} computes the second solutions at $N=4$ reported in
Appendix~\ref{app:sim-main}.

\paragraph{The boundary $\rho = 1$.}
In the limit $\rho \to 1$ ($s \to 0$) the beta distributions degenerate to point
masses at 0 and 1, so all $N$ ratings of an item coincide and $k_c \in \{0, N\}$
with $\Pr(k_c = N) = (1-\theta)\eps_0 + \theta(1-\eps_1)$. The distribution of
$k_c$ then depends on three parameters through a single quantity, and the
parameters are not identifiable for any $N$. For $\rho$ below one and $N \ge 5$
the model is identified for almost every parameter value, but the information
that separates the parameters diminishes as $\rho$ approaches one. At $N = 4$ the
loss is exact for some parameter values. For the parameter used in
Section~\ref{sec:sims}, $\theta = 0.3$, $\eps_0 = 0.15$ and $\eps_1 = 0.10$, the
system at $N = 4$ has a second admissible solution when $\rho = 0.75$ and when
$\rho = 0.90$, and in both cases that solution has lower error rates and a lower
correlation than the truth. At $\rho = 0.75$ it is $\theta' = 0.362$,
$\eps_0' = 0.065$, $\eps_1' = 0.079$ and $\rho' = 0.555$. We regard this as a
heuristic explanation for the direction of the bias observed at $N \ge 5$ in
Section~\ref{sec:sims}, not as a proof of it.

\subsection{Model extensions}
\label{app:ext}

\paragraph{Covariates and treatment effects.}
Let $Z_c = (X_c, T_c, H_c)$ collect the item covariates, the treatment indicator,
and the difficulty attribute. The extended model specifies
\begin{align*}
\mathrm{logit}(\theta_c) &= \alpha_\theta + X_c\beta + \tau T_c, \\
\eps_{0c} &= u_c v_c, \qquad \eps_{1c} = (1-u_c) v_c, \\
\mathrm{logit}(u_c) &= \alpha_u + \gamma_u H_c, \\
\mathrm{logit}(v_c) &= \alpha_v + \gamma_v H_c,
\end{align*}
with a common correlation $\rho$, and $k_c$ given $Z_c$ follows
Equation~\eqref{eq:bb} with the item-specific parameters. The restriction
$\eps_{0c}+\eps_{1c} = v_c < 1$ holds for every item by construction. The error
rates depend on $H_c$ but not on $T_c$, so measurement error is assumed not to
differ between treatment arms. Under consistency, $D_c = D_c(T_c)$, ignorability of
treatment given the covariates, $(D_c(1), D_c(0)) \perp T_c \mid X_c$, and
positivity, $0 < \Pr(T_c = 1 \mid X_c) < 1$, the average treatment effect is
$\eta = E[\mathrm{expit}(\alpha_\theta + X_c\beta + \tau) - \mathrm{expit}(\alpha_\theta + X_c\beta)]$,
where the expectation is over the distribution of $X_c$.

\begin{theoremA}[Identifiability with covariates]
\label{thm:glm}
Assume $N \ge 5$ and $0 \le \rho < 1$. Assume also that, over the support of
$Z_c$, the design matrix with rows $(1, X_c, T_c)$ and the design matrix with
rows $(1, H_c)$ both have full column rank. Then the parameters
$(\alpha_\theta, \beta, \tau, \alpha_u, \gamma_u, \alpha_v, \gamma_v, \rho)$ are
identifiable for almost every value of these parameters, and so is $\eta$ under
consistency, ignorability of treatment given $X_c$, and positivity.
\end{theoremA}

\begin{proof}
Choose finitely many support points $z_1, \dots, z_K$ of $Z_c$ at which both
design matrices have full column rank. At each $z_k$ the conditional distribution
of $k_c$ is Equation~\eqref{eq:bb} with parameter
$\psi_k = (\theta(z_k), \eps_0(z_k), \eps_1(z_k), \rho)$. The map from the
regression parameters to $\psi_k$ is continuously differentiable and has full
rank four, since $\alpha_\theta$, $\alpha_u$, $\alpha_v$ and $\rho$ move the four
components of $\psi_k$ independently and $(u,v) \mapsto (uv, (1-u)v)$ has
Jacobian determinant $v > 0$. The preimage of the null set $\mathcal{Z}$ of
Theorem~\ref{thm:bb}(iii) under such a map is null, and so is the finite union
over $k$. Outside this union, Theorem~\ref{thm:bb} identifies each $\psi_k$, and
therefore $\mathrm{logit}\,\theta(z_k)$, $\mathrm{logit}\,u(z_k)$ and
$\mathrm{logit}\,v(z_k)$ for $k = 1,\dots,K$. These are linear in the regression
coefficients, and full column rank identifies the coefficients uniquely. Under consistency, ignorability of treatment given $X_c$, and positivity, the
average treatment effect $\eta$ is a functional of identified quantities.
\end{proof}

\paragraph{Semi-supervised model.}
Let $L_c = 1$ if the true state of item $c$ is revealed. Labeled items
contribute $\theta_c^{D_c}(1-\theta_c)^{1-D_c}\Pr(k_c\mid D_c, Z_c)$ to the
likelihood and unlabeled items contribute Equation~\eqref{eq:bb}. In the
semi-supervised model the error rates are parameterized directly as
$\mathrm{logit}(\eps_{0c}) = \alpha_0 + \gamma_0 H_c$ and
$\mathrm{logit}(\eps_{1c}) = \alpha_1 + \gamma_1 H_c$, without the restriction
$\eps_{0c}+\eps_{1c}<1$. The logit links keep $\theta_c$, $\eps_{0c}$ and
$\eps_{1c}$ in $(0,1)$.

\begin{theoremA}[Identifiability of the semi-supervised model]
\label{thm:semisup}
Assume $N \ge 2$, $0 \le \rho < 1$, that $\theta_c$, $\eps_{0c}$ and $\eps_{1c}$
lie in $(0,1)$, that $L_c$ is independent of $(D_c, k_c)$ given $Z_c$ with
$\Pr(L_c = 1 \mid Z_c) > 0$ on the support of $Z_c$, and the rank conditions of
Theorem~\ref{thm:glm} with $(1, H_c)$ as the design for $\mathrm{logit}(\eps_{0c})$
and $\mathrm{logit}(\eps_{1c})$. Then all parameters are identifiable for
every parameter value, with no restriction on the sign of $J$.
\end{theoremA}

\begin{proof}
Among labeled items with $Z_c = z$ and $D_c = d$, $k_c$ follows a single
beta-binomial distribution with mean propensity $\mu_d(z)$ and correlation
$\rho$. Its first two moments are $m_1 = \mu_d(z)$ and
$m_2 = \mu_d(z)(\mu_d(z) + \kappa)/(1+\kappa)$ with $\kappa = \rho/(1-\rho)$, so
$\kappa = (m_2 - m_1^2)/(m_1 - m_2)$, where $m_1 - m_2 = E[\pi(1-\pi)] > 0$ whenever
$0 < \mu_d(z) < 1$. Hence $\eps_0(z) = \mu_0(z)$, $\eps_1(z) = 1-\mu_1(z)$, and
$\rho$ are identified when $N \ge 2$. Because $L_c$ is independent of $D_c$ given
$Z_c$, $\theta(z) = \Pr(D_c = 1 \mid Z_c = z, L_c = 1)$ is identified from the
labeled items. The regression coefficients then follow from full column rank as
in Theorem~\ref{thm:glm}. No step uses the sign of $J$.
\end{proof}

Without labels, a model with $J < 0$ is not identifiable. In the model without
covariates, the relabeled parameter $(1-\theta, 1-\eps_1, 1-\eps_0, \rho)$
produces the same distribution of $k_c$ and has acceptance gap $-J > 0$. When
$N \ge 5$ and the relabeled parameter lies outside the exceptional set of
Theorem~\ref{thm:bb}(iii), the model restricted to $J > 0$ therefore targets the
relabeled parameter, and under standard regularity conditions its estimate of
the prevalence converges to $1-\theta$ rather than $\theta$. With covariates on
the error rates the relabeled error rates need not follow the assumed
logit-linear form, and if $J$ changes sign across items there is no single
relabeled parameter, so the restricted model is then misspecified and its limit
is a best-fitting approximation. When $J = 0$ the two components coincide and the
prevalence is not identifiable. Theorem~\ref{thm:semisup} shows that a labeled
subset removes these difficulties.

\paragraph{Weighting of labeled items.}
The labeled block may be upweighted by a factor $W$ that multiplies its
log-likelihood. Setting $W=1$ yields an ordinary posterior. Setting $W>1$ raises
the labeled likelihood to a power greater than one, which is not a Bayesian
posterior for any generative model. Section~\ref{sec:anchors} and
Appendix~\ref{app:weight} show that $W > 1$ does not improve accuracy on the
benchmark and narrows intervals, and we therefore recommend $W=1$. The
semi-supervised fits in Table~\ref{tab:twobytwo} were run with
$W = W_{\text{eq}}$, and the labeling experiment reports $W = 1$.

\section{Details of the empirical study}
\label{app:empirical}

\subsection{Prompt conditions}
\label{app:prompts}

All three prompt conditions are applied to the same 250 items, so that
differences between them reflect the prompt rather than the sample.

\paragraph{Arm A.}
The judge prompt published by \citet{yan2026verifybench}, reproduced verbatim. It
supplies the question, the reference answer, and the response, instructs the
judge to compare final results rather than reasoning and to treat equivalent
answers as consistent, and asks for a final line stating the judgment. Using the
prompt without modification makes our error rates comparable with those reported
by the benchmark authors.

\paragraph{Arm B.}
Arm~A with one additional instruction asking the judge to report the difficulty
of the verification task on a scale from 1 to 5. Running Arms~A and~B on the same
items allows the effect of eliciting the rating to be measured. For the two
largest models, 1.2\% and 2.0\% of items change their count $k_c$ by more than two
between the arms. For the two smallest models the proportions are 22.1\% and
20.8\%.

\paragraph{Arm C.}
The prompt published by \citet{yan2026verifybench} for the setting without a
reference answer. It supplies the question and the response, and the judge must
determine the correct answer itself before deciding whether the response matches
it. The benchmark authors report an accuracy cost of 5 to 18 percentage points
for withholding the reference.

\subsection{Sampling and parse failures}
\label{app:task}

We sample 150 items from VerifyBench and 100 from VerifyBench-Hard with a fixed
seed. The realized prevalence is $0.493$ in the VerifyBench sample against
$0.500$ in the full split, and $0.240$ in the VerifyBench-Hard sample against
$0.291$ in the full split. The deviation on VerifyBench-Hard is about $1.1$
standard errors for a sample of 100 items.

Of the 52{,}500 judgments, 94.2\% yielded a parseable verdict. Failures are
concentrated in the smallest model, which produced unparseable output in 23\% of
ratings, with a mean output length of 131 characters against 738 to 3324 for the
other models. Recomputing the total error rate with unparsed ratings counted as
positive, counted as negative, or dropped changes it by at most $0.061$ across
models and by less than $0.02$ for five of the seven.

After unparsed ratings are dropped, an item may have fewer than ten ratings.
Items with fewer than three parseable ratings are excluded from the fits,
because Theorem~\ref{thm:binom} requires $N_c \ge 3$. This retains 5{,}046 of the
5{,}250 combinations of item and setting. Exclusions are concentrated in
gemma-3-1b-it under Arm~B, which retains 79 of 150 items on VerifyBench and 52 of
100 on VerifyBench-Hard. Across the other settings at most 13 items are excluded.
Retained items enter the likelihood with their own $N_c$. The identification
results of Theorems~\ref{thm:binom} and~\ref{thm:bb} are stated for a common $N$
and apply to the items that share it.

The responses in the sample were generated by 22 distinct models. One of these,
gemma-3-1b-it, is also a judge in our study and generated 26 of the 250
responses, 15 on VerifyBench and 11 on VerifyBench-Hard. None of the other six
judges generated any response in the sample.

\subsection{Computing the design effect}
\label{app:de-method}

For a group of items with the same true state, let $\hat p$ be the mean acceptance
fraction $k_c/N$. Under conditional independence
$\mathrm{Var}(k_c/N) = p(1-p)/N$. The design effect is the ratio
\[
\DE \;=\; \frac{\widehat{\mathrm{Var}}(k_c/N)}{\hat p(1-\hat p)/N}
\;=\; 1+(N-1)\rho,
\]
and $\Neff = N/\DE$. When $N_c$ varies across items we use the observed
fractions $k_c/N_c$ and set $N$ to the median of $N_c$ within the group. The
ratio is truncated below at one. We compute it separately for items with correct
and incorrect responses and pool the two by sample size. Computing it on the pooled
sample would confound correlation within items with the difference in acceptance
rates between the two groups, and would report a large design effect for an
accurate model that simply answers correctly in both groups. On the pooled sample
the two models that reason by default appear to have much higher correlation than
the others, and this difference disappears once the computation conditions on the
true state.

\subsection{The difficulty covariate}
\label{app:difficulty}

The difficulty rating reported under Arm~B is weakly and positively associated
with realized error. The Spearman correlation between the mean reported
difficulty of an item and the fraction of its ratings that disagree with the
human label lies between $0.18$ and $0.31$ across models, and the error rate
increases monotonically across reported difficulty levels for six of the seven
models. Conditioning on the reported difficulty nonetheless removes between 0\%
and 35\% of the excess variance, and less than 15\% for four of the six models for
which the comparison is available. Part of the explanation is range compression.
The models rate almost every item as easy, with one placing all 250 items at
levels one or two and another placing 181 of 244 at level one. A covariate on the
error rates therefore does not remove the need for the beta-binomial likelihood.

\subsection{Stratified selection of labeled items}
\label{app:strat}

In the experiments of Section~\ref{sec:anchors}, labeled items are drawn
stratified on the true state, allocated in proportion to the class sizes with at
least one item per class. In the VerifyBench-Hard sample, where 24 of the 100
responses are correct, a simple random sample of five items contains no correct
response with probability $0.245$ and no incorrect response with probability
$0.001$. In those cases the labeled subset alone cannot determine the labeling of
the two components. Stratification avoids this, but proportional allocation also
makes the prevalence among labeled items close to the sample prevalence. Labeled
selection then depends on the true state, which departs from the condition of
Theorem~\ref{thm:semisup}, and the reductions in prevalence error at small
labeled fractions in Tables~\ref{tab:twobytwo} and~\ref{tab:anchors} are larger
than a simple random sample of the same size would give. In applications where
the true state is unknown before labeling, a simple random sample of a
sufficiently large subset is required, and Appendix~\ref{app:sim-semisup} uses
one.

\subsection{Fitting and convergence}
\label{app:fitting}

Each model is fitted with four chains of 2000 iterations, half of them warmup. A
fit that does not meet the convergence criteria is refitted once with 6000
iterations. In preliminary runs, raising the target acceptance rate from $0.95$ to
$0.99$ did not resolve convergence failures, whereas tripling the number of
iterations did, which indicates a mixing problem rather than a step size problem.
A fit is accepted when $\widehat{R}<1.01$, the smallest effective sample size
exceeds 400, and there are no divergent transitions. These criteria are applied
to model parameters only. The posterior probabilities $\Pr(D_c=1\mid k_c)$ and the
pointwise log likelihood are excluded, because for an item whose posterior
probability is nearly constant across draws the effective sample size approaches
zero for reasons unrelated to the exploration of the posterior. The 182 fits comprise
42 settings for each of four models, namely the binomial model with each error
rate bounded below $1/2$, the binomial model of Equation~\eqref{eq:binom} with the
restriction of Equation~\eqref{eq:uv}, the beta-binomial model of
Equation~\eqref{eq:bb}, and the semi-supervised beta-binomial model with 10\% of
items labeled, together with 14 fits of the model with the difficulty covariate
to the Arm~B settings. The first of these is not reported in the main text. Its
coverage of the false acceptance rate is $0.000$ on both splits, as for the
binomial model in Table~\ref{tab:coverage}. Of the 182 fits, 181 met the criteria,
14 of them after the longer run, and one did not converge after the longer run. The labeling experiment of
Section~\ref{sec:anchors} comprises a further 684 fits, which are not included in
this count.

\subsection{Coverage on VerifyBench-Hard}
\label{app:vbh}

On VerifyBench-Hard the beta-binomial interval covers the false acceptance rate
in $0.571$ of settings and the false rejection rate in $0.667$, against $0.905$
and $1.000$ on VerifyBench. The seven VerifyBench-Hard settings with $J \le 0$,
all under Arm~C, have acceptance gaps between $-0.32$ and $-0.10$. In these
settings the identifying condition fails, the model restricted to $J>0$ cannot
represent the true parameter, and its intervals miss both error rates in every
case. Excluding them, coverage on VerifyBench-Hard is $12/14 = 0.857$ for
$\eps_0$ and $14/14 = 1.000$ for $\eps_1$.

The four remaining misses, two on each split, all concern the false acceptance
rate of gemma-3-1b-it. They occur under Arm~B and Arm~C on VerifyBench, with
$J = 0.083$ and $0.158$, and under Arm~A and Arm~B on VerifyBench-Hard, with
$J = 0.066$ and $0.108$. The intraclass correlation in these settings lies
between $0.78$ and $0.96$. These are the settings in which Section~\ref{sec:separation}
finds recovery weakest and in which Theorem~\ref{thm:bb} and
Section~\ref{sec:sims} indicate that the information separating the parameters
is smallest.

Imprecision of the reference values plays a smaller role than these two
factors. The false rejection rate on VerifyBench-Hard is computed from 10 to 24
items with a correct response and has a standard error near $0.09$, but its
coverage is complete in every setting with $J>0$. The false acceptance rate is
computed from 42 to 76 items with an incorrect response, with a standard error of
at most $0.08$.

\subsection{Weight on labeled items}
\label{app:weight}

Across the labeling experiment, the mean absolute error in the estimated
prevalence is $0.022$, $0.020$ and $0.022$ for $W=1$, $W=3$ and
$W=W_{\text{eq}}$, while the mean interval width is $0.292$, $0.190$ and $0.170$.
For comparison, the unsupervised fit on the same settings has a mean interval
width of $0.630$ and a mean absolute error of $0.110$. Upweighting narrows the
intervals by a factor of $1.7$ without improving accuracy. With 5\% of items labeled in the settings with an acceptance gap of about
$0.62$, the error rises from $0.025$ to $0.040$ at $W=1$. This increase does not
persist at larger labeled fractions, and we attribute it to variation in which
items are drawn, since each entry averages five draws of the labeled subset in
each of three settings. At labeled fractions of 35\% and 50\%,
$W_{\text{eq}} = 1$, so $W=1$ and $W = W_{\text{eq}}$ coincide there.

\section{Details of the simulation studies}
\label{app:sims}

\subsection{Main simulation study}
\label{app:sim-main}

For each item, $D_c \sim \mathrm{Bernoulli}(\theta)$, the acceptance propensity is
drawn as $\pi_c \sim \mathrm{Beta}(p s, (1-p) s)$ with $p = \eps_0$ if $D_c = 0$
and $p = 1-\eps_1$ if $D_c = 1$ and $s = (1-\rho)/\rho$, and the $N$ ratings are
independent $\mathrm{Bernoulli}(\pi_c)$ draws given $\pi_c$. At $\rho = 0$ the
propensity equals $p$. Both latent class models are fitted by maximum likelihood
with the L-BFGS-B algorithm from several starting values, with all coordinates on
the logit scale, so that the restriction $J>0$ holds by construction and
relabeling cannot occur. The likelihood is evaluated on the histogram of counts,
which is a sufficient statistic. Intervals are Wald intervals formed on the
unconstrained coordinates from the numerically differentiated observed
information and mapped through the inverse logit. Because $\theta$ and
$\eps_0+\eps_1 = v$ are each a monotone function of a single coordinate, the
mapping is exact. Replicates in which the observed information is not positive
definite contribute to the bias but not to the coverage. For the beta-binomial
model this occurred in 123 to 130 of 250 replicates at $N = 3$ and in at most 37
of 250 replicates otherwise. Majority vote breaks ties at random when $N$ is even.

For the parameter used here, the moment equations at $N = 4$ have a second
admissible solution when $\rho = 0.75$, namely $\theta' = 0.362$,
$\eps_0' = 0.065$, $\eps_1' = 0.079$ and $\rho' = 0.555$, and when $\rho = 0.90$,
namely $\theta' = 0.374$, $\eps_0' = 0.021$, $\eps_1' = 0.031$ and $\rho' = 0.516$.
No second admissible solution exists at $\rho \in \{0.25, 0.50\}$.

\begin{table*}[t]
\centering\small
\setlength{\tabcolsep}{4.5pt}
\caption{Bias and coverage for the total error rate at $N \in \{5, 10, 20\}$. Settings as in Table~\ref{tab:sim-total}.}
\label{tab:sim-total-full}
\begin{tabular}{llrrrrrr}
\toprule
 & & \multicolumn{2}{c}{$N=5$} & \multicolumn{2}{c}{$N=10$} & \multicolumn{2}{c}{$N=20$} \\
\cmidrule(lr){3-4} \cmidrule(lr){5-6} \cmidrule(lr){7-8}
$\rho$ & Likelihood & Bias & Cover & Bias & Cover & Bias & Cover \\
\midrule
0.00 & Binomial & $-0.001$ & 0.94 & $-0.002$ & 0.96 & $0.000$ & 0.94 \\
 & Beta-binomial & $0.000$ & 0.96 & $-0.002$ & 0.96 & $0.000$ & 0.94 \\
\addlinespace[2pt]
0.25 & Binomial & $-0.039$ & 0.48 & $-0.021$ & 0.59 & $-0.018$ & 0.49 \\
 & Beta-binomial & $-0.004$ & 0.93 & $+0.004$ & 0.95 & $-0.001$ & 0.97 \\
\addlinespace[2pt]
0.50 & Binomial & $-0.104$ & 0.02 & $-0.087$ & 0.01 & $-0.083$ & 0.00 \\
 & Beta-binomial & $-0.032$ & 0.71 & $-0.016$ & 0.88 & $-0.022$ & 0.84 \\
\addlinespace[2pt]
0.75 & Binomial & $-0.177$ & 0.00 & $-0.168$ & 0.00 & $-0.163$ & 0.00 \\
 & Beta-binomial & $-0.115$ & 0.58 & $-0.100$ & 0.56 & $-0.101$ & 0.58 \\
\addlinespace[2pt]
0.90 & Binomial & $-0.221$ & 0.00 & $-0.217$ & 0.00 & $-0.216$ & 0.00 \\
 & Beta-binomial & $-0.199$ & 0.37 & $-0.187$ & 0.36 & $-0.187$ & 0.39 \\
\bottomrule
\end{tabular}
\end{table*}

\subsection{Semi-supervised model and prediction-powered inference}
\label{app:sim-semisup}

Data are generated from Equation~\eqref{eq:bb} with $C = 500$ items, $N = 10$,
$\rho = 0.5$ and $\theta = 0.3$, in three settings of the error rates. With
$\eps_0 = 0.15$ and $\eps_1 = 0.10$ the acceptance gap is $J = 0.75$ and the
unsupervised model is identified. With $\eps_0 = 0.45$ and $\eps_1 = 0.35$ the gap
is $J = 0.20$ and the model is identified but the classes are weakly separated.
With $\eps_0 = 0.70$ and $\eps_1 = 0.60$ the gap is $J = -0.30$, the LLM is
systematically wrong, and the unsupervised model is not identified. In each
replicate the true state of a simple random sample of 50 items is revealed.

Five procedures are compared. The semi-supervised beta-binomial model is fitted
by maximum likelihood without the restriction $J>0$ and with $W = 1$. The
unsupervised beta-binomial model uses all items and ignores the labels.
Prediction-powered inference \citep{angelopoulos2023prediction} uses the majority
label as the prediction, averages it over the unlabeled items, and adds the mean
difference between the true state and the prediction over the labeled items, with
the interval given by \citet{angelopoulos2023prediction}. The labeled mean uses
the 50 labeled items alone. Majority vote uses the proportion of positive
majority labels over all items.

\begin{table*}[t]
\centering\small
\setlength{\tabcolsep}{4pt}
\caption{Estimation of the prevalence ($\theta = 0.3$) with a 10\% labeled subset, $C = 500$ items, $N = 10$ ratings and $\rho = 0.5$, over 300 replicates. RMSE is the root mean squared error, Cover the coverage of nominal 95\% intervals, and AUC the ranking accuracy on unlabeled items. Prediction-powered inference and the labeled mean return no probabilities for individual items.}
\label{tab:sim-semisup}
\begin{tabular}{lrrrrrrrrr}
\toprule
 & \multicolumn{3}{c}{$J = 0.75$} & \multicolumn{3}{c}{$J = 0.20$} & \multicolumn{3}{c}{$J = -0.30$} \\
\cmidrule(lr){2-4} \cmidrule(lr){5-7} \cmidrule(lr){8-10}
Procedure & RMSE & Cover & AUC & RMSE & Cover & AUC & RMSE & Cover & AUC \\
\midrule
Semi-supervised & 0.037 & 0.95 & 0.96 & 0.064 & 0.94 & 0.64 & 0.059 & 0.95 & 0.72 \\
Prediction-powered inference & 0.048 & 0.95 & n/a & 0.089 & 0.94 & n/a & 0.108 & 0.95 & n/a \\
Labeled items only & 0.068 & 0.92 & n/a & 0.065 & 0.93 & n/a & 0.063 & 0.94 & n/a \\
Unsupervised & 0.083 & 0.67 & 0.96 & 0.318 & 0.33 & 0.65 & 0.375 & 0.32 & 0.27 \\
Majority vote & 0.067 & 0.14 & 0.96 & 0.215 & 0.00 & 0.65 & 0.321 & 0.00 & 0.27 \\
\bottomrule
\end{tabular}
\end{table*}

Table~\ref{tab:sim-semisup} reports the results. The semi-supervised model and
prediction-powered inference are both approximately unbiased, with biases no
larger than $0.01$ in absolute value, and both attain coverage between $0.94$ and
$0.95$ in every setting. The semi-supervised model has the smaller root mean
squared error and the narrower interval, with mean widths of $0.14$, $0.24$ and
$0.23$ against $0.19$, $0.34$ and $0.43$ for prediction-powered inference. This
advantage reflects the fact that the data are generated from the model, and
prediction-powered inference retains its validity without that assumption. Its
efficiency also depends on the quality of the prediction. When $J<0$ the majority
label is negatively associated with the true state, and the rectifier absorbs
most of the variation. When $J = 0.20$ or $J = -0.30$, estimation from the
labeled items alone is nearly as accurate as the semi-supervised model, which
indicates that the ratings add little information about the prevalence when the
classes are weakly separated.

Without labels, the beta-binomial model overestimates the prevalence by $0.21$
when $J = 0.20$ and by $0.30$ when $J = -0.30$, and majority vote is biased in
every setting. When $J = -0.30$ the restricted model targets the relabeled
prevalence $1-\theta = 0.70$, which differs from the truth by $0.40$. The average
bias of $0.30$ is smaller because the estimate does not reach the relabeled
value in every replicate, reflecting weak identification at $|J| = 0.30$ and
$\rho = 0.5$ with 500 items. When $J = -0.30$ the unsupervised model ranks items in the reverse
order, with an AUC of $0.27$, and the semi-supervised model restores the correct
direction, with an AUC of $0.72$. When $J > 0$ the labels do not improve the AUC. The
posterior probability is then an increasing function of $k_c$ under both models,
so the two rankings coincide, except in replicates in which the semi-supervised
fit estimates the direction of separation incorrectly. Such replicates lower the
AUC from $0.654$ to $0.641$ at $J = 0.20$.

\section{Broader impact, reproducibility and compute}
\label{app:repro}

\paragraph{Broader impact.}
This work is diagnostic. It measures how much uncertainty an established
aggregation practice conceals. We expect the main effect of adopting its
recommendations to be wider and better calibrated intervals on quantities
estimated from LLM annotations, together with a more explicit accounting of when
gold labels are needed. A possible negative effect is that practitioners treat an
adequate acceptance gap as a reason to deploy a judge without labeled
validation. Estimating the gap itself requires labels.

\paragraph{Reproducibility.}
Model checkpoints are pinned to commit hashes and served at their released
precision. Items are sampled with a fixed seed, and because item identifiers
reference rows of the source benchmark, a larger sample drawn with the same seed
contains the one reported here. Prompts are reproduced verbatim from the benchmark
authors' appendices, and the predictions in Table~\ref{tab:judges} were fixed
before data collection. We release, at
\url{https://github.com/google-research/ljer}, the collected ratings keyed to
benchmark item identifiers, together with a retrieval script, the
Stan models, the simulation code, and the scripts for the exact computations in
Appendix~\ref{app:proof-bb}. Source text from the benchmark is not redistributed
and is obtained from its original repository under its own license. The ratings
record, for each
rating, the model, the checkpoint revision, the prompt condition, the round, the
verdict, and the reported difficulty where elicited.

\paragraph{Compute.}
Rating collection used one NVIDIA A100 80GB GPU for approximately two hours,
serving each model in turn with vLLM. Model fitting ran on a laptop CPU and
completed the 182 fits, each with four chains, in about fifteen minutes. The
simulation studies run on two CPU cores in under an hour.

\end{document}